\documentclass[%
 aip,
 amsmath,amssymb,
 reprint,%
]{revtex4-1}

\usepackage{graphicx}
\usepackage{dcolumn}
\usepackage{bm}

\usepackage[utf8]{inputenc}
\usepackage[T1]{fontenc}
\usepackage{mathptmx}
\usepackage{etoolbox}
\usepackage{xspace}
\usepackage{gensymb}
\usepackage{verbatim}
\usepackage{float}
\usepackage{xcolor}
\usepackage{enumitem}
\usepackage{placeins}
\makeatletter
\def\@email#1#2{%
 \endgroup
 \patchcmd{\titleblock@produce}
  {\frontmatter@RRAPformat}
  {\frontmatter@RRAPformat{\produce@RRAP{*#1\href{mailto:#2}{#2}}}\frontmatter@RRAPformat}
  {}{}
}%
\makeatother
\begin{document}

\preprint{AIP/123-QED}

\title[]{Effect of leading edge geometry on aerodynamic performance of a flat plate }
\author{Intesaaf Ashraf}
 \altaffiliation[]{Mechanical Engineering Department, University College London, WC1E 7JE, London,  United Kingdom}
 \email{intesaaf.ashraf@ucl.ac.uk}
\author{Hamid Hassan Khan}%
 \email{hamidhassan.k@srmap.edu.in}
\altaffiliation[]{Department of Mechanical Engineering, SRM University-AP, Amaravati 522240, Andhra Pradesh, India}%
 
\author{Majid Hassan Khan}%
 \email{mhkhan@dubai.bits-pilani.ac.in}
\altaffiliation[]{Department of Mechanical Engineering, Birla Institute of Technology and Science, Pilani, Dubai Campus, Dubai International Academic City, Dubai, United Arab Emirates}%

\date{\today}

\begin{abstract}
Flow separation around flat plate is influenced by the thickness and leading edge geometry, which creates the shear stresses and pressure distribution manifesting into unsteady aerodynamic forces. Leading edge geometries have been widely reported but the influence of leading edge curvature at low Reynolds numbers (Re) have found less attention. The present study uses two dimensional numerical simulation to investigate the effect of a hemispherical curvature (protuberance) at the leading edge and reports how vortex shedding, shedding frequencies and aerodynamic forces are affected over a range of angles of attack ($0^\circ \leq \alpha \leq 25^\circ$) at Re = 1000.  Thicker vortical regions are formed near the plate with curvature at the leading edge. Onset of shedding is noted at angle of attack of $5^\circ$ for the plate with curvature at the leading edge and it gets delayed  for simple flat plate. The temporal force coefficients on the plates show multiple frequencies at higher angles of attack. The plate with curvature at the leading edge has a glide ratio of 1 at $0^\circ$ angle of attack and has equivalent glide ratio to that of simple flat plate at higher angles of attack.  The simple flat plate has better glide ratios at moderate angles of attack like $5^\circ, 10^\circ$ and $15^\circ$.  The results will help in making engineering informed design decisions for smaller unmanned aerial/aquatic vehicles cruising at low Reynolds numbers. 

\end{abstract}

\maketitle

\section{Introduction}

The leading edge is the most important region of a lifting surface. The incoming flow meets the body there first, divides at the stagnation point and accelerates around the nose. On the suction side of a wing at incidence, the suction peak forms within the first few per cent of the chord, and the strongest adverse pressure gradient acts just behind it. The boundary layer is thinnest in this region and separates easily. What happens near the leading edge therefore controls the maximum lift and the type of stall \citep{smith1975}. The nose radius is one of the main parameters in airfoil design, and high-lift devices such as slats and Krueger flaps act on the leading edge \citep{smith1975}. Small changes in this region, caused by roughness, insects or ice, can produce large losses in performance.

The link between nose geometry and stall has been known for a long time. \citet{mccullough1951} identified three types of stall: trailing-edge stall, leading-edge stall and thin-airfoil stall. \citet{gault1957} showed that the type of stall depends mainly on the upper-surface ordinate at $1.25\%$ chord and on the Reynolds number. The mechanism behind leading-edge and thin-airfoil stall is the laminar separation bubble. The laminar boundary layer separates under the adverse pressure gradient, the separated shear layer becomes turbulent, and the flow reattaches to the surface. The result is a short bubble, which can grow and burst into a long bubble as the angle of attack increases \citep{gaster1967,horton1968,tani1964}. These bubbles are also unsteady: the shear layer sheds vortices near reattachment, and the bubble as a whole moves back and forth at a low frequency \citep{kiya1983,cherry1984,yarusevych2009}.

The flat plate is the simplest geometry in which these effects can be studied. It has the sharpest possible leading edge and no thickness or camber, so the flow is governed by the nose alone. At zero incidence it reduces to the classical boundary layer: the Blasius solution with $C_f = 0.664\,Re_x^{-1/2}$ in the laminar region \citep{prandtl1905,blasius1907}, and correlations such as $C_f \approx 0.0592\,Re_x^{-1/5}$ after transition to turbulence \citep{Schlichting2017}. At any positive angle of attack the flow separates at the leading edge itself \citep{fage1927}. The separated shear layer reattaches downstream and closes a bubble, which lengthens as the angle of attack increases until the plate stalls in the thin-airfoil manner \citep{crompton2000}. Flat plates are also of practical interest. At chord Reynolds numbers below about $10^5$ they perform as well as or better than conventional airfoils \citep{pelletier2000,winslow2018}, which is why they are used on micro air vehicles and small drones \citep{mueller2003}, and their aerodynamic forces have been measured for many planforms and aspect ratios \citep{taira2009,ananda2015}.

Because the leading edge governs the flow, modifying it is a common method of passive flow control. The best-known example is the tubercled leading edge, inspired by the flipper of the humpback whale \citep{fish1995}. Sinusoidal protuberances along the span delay stall and make the loss of lift after stall more gradual \citep{johari2007,hansen2011}. The leading-edge shape can also change unintentionally. Ice that forms as horns or ridges near the nose causes strong separation and large performance losses \citep{bragg2005,lee2003}. These studies deal with three-dimensional tubercles or with irregular ice shapes.  \textcolor{black}{Earlier studies have investigated leading edge effects using (i) sharp leading edge \citep{crompton2000,fujiwara2020experimental,Smith2021,joshi2025hydrodynamics}, (ii) rectangular or blunt  leading edge \citep{cherry1984,ota1976separated,ota1981reattachment, kiya1983}, (iii) round semi-circular leading edge \citep{yang2001large,samson2016effects, anand2017features, anand2022characteristics,stevenson2016particle}  and (iv) optimized leading edge e.g., elliptical \citep{wanderley2001boundary,schrader2010receptivity, hanson2012aerodynamic}. \citet{ota1981reattachment} investigated a flat plate with different modified leading edges at low Re $\in$ [40, 2000]. Using laser anemometry techniques \citet{crompton2000} investigated flow around a $20^\circ$ chamfered flat plate with thickness 3.75\% of the length at various incidence angle. They reported the bubble length dependence on angle of attack becoming weak at $Re \approx 10^5$ and secondary recirculation bubble near the leading edge.  \citet{Smith2021} provided a reattachment model for a flat plate at small angle of attach at $Re \approx 10^5$. \citet{cherry1984} used experimentally measured pressures and performed smoke flow visualization at $Re \approx 3.2 \times 10^4 \pm 0.2 \times 10^4$ and reported the separated and reattachment flow at $0^\circ$ angle of attack. They  observed low frequency unsteadiness superimposed on the base vortical shedding revealing several vortex trains. Similar work with blunt $90^\circ$ cornered flat plates was reported by \citet{kiya1983} and \citet{ota1976separated} reports the unsteadiness in shedding at $0^\circ$ at $Re \approx O(10^4)$  and presence of vortical structures in the reattachment bubbles. \citet{yang2001large} also reported large vortical structures reattaching for a flow around flat plate with semi-circular leading edge at $0^\circ$ angle of attach. \citet{stevenson2016particle} reported shortening of the recirculation bubble by almost three times when the free-stream turbulence increases to $7\%$ for flow around a flat plate with semi-circular leading edge at $Re = 800$ (based on leading edge diameter) . Further refinements were done in the leading edge geometry by using elliptical leading edge \citep{schrader2010receptivity,wanderley2001boundary,hanson2012aerodynamic}. \citet{hanson2012aerodynamic} investigated varying elliptical geometries to develop an optimized asymmetric modified super ellipse leading edge which reduces non-zero and adverse pressure gradient regions. }

\textcolor{black}{Earlier works focused on re-shaping the leading edge and reported the flow behavior around the flat plate. There is a dearth of literature investigating the effect of protuberance near a flat blunt $90^\circ$ leading edge. Similarly the engineering parameters like Strouhal numbers and force coefficients has found little attention  for flat plate with}  a single, simple, well-defined protuberance at the leading edge.  

\begin{figure}
    \centering
    \includegraphics[width=1\linewidth]{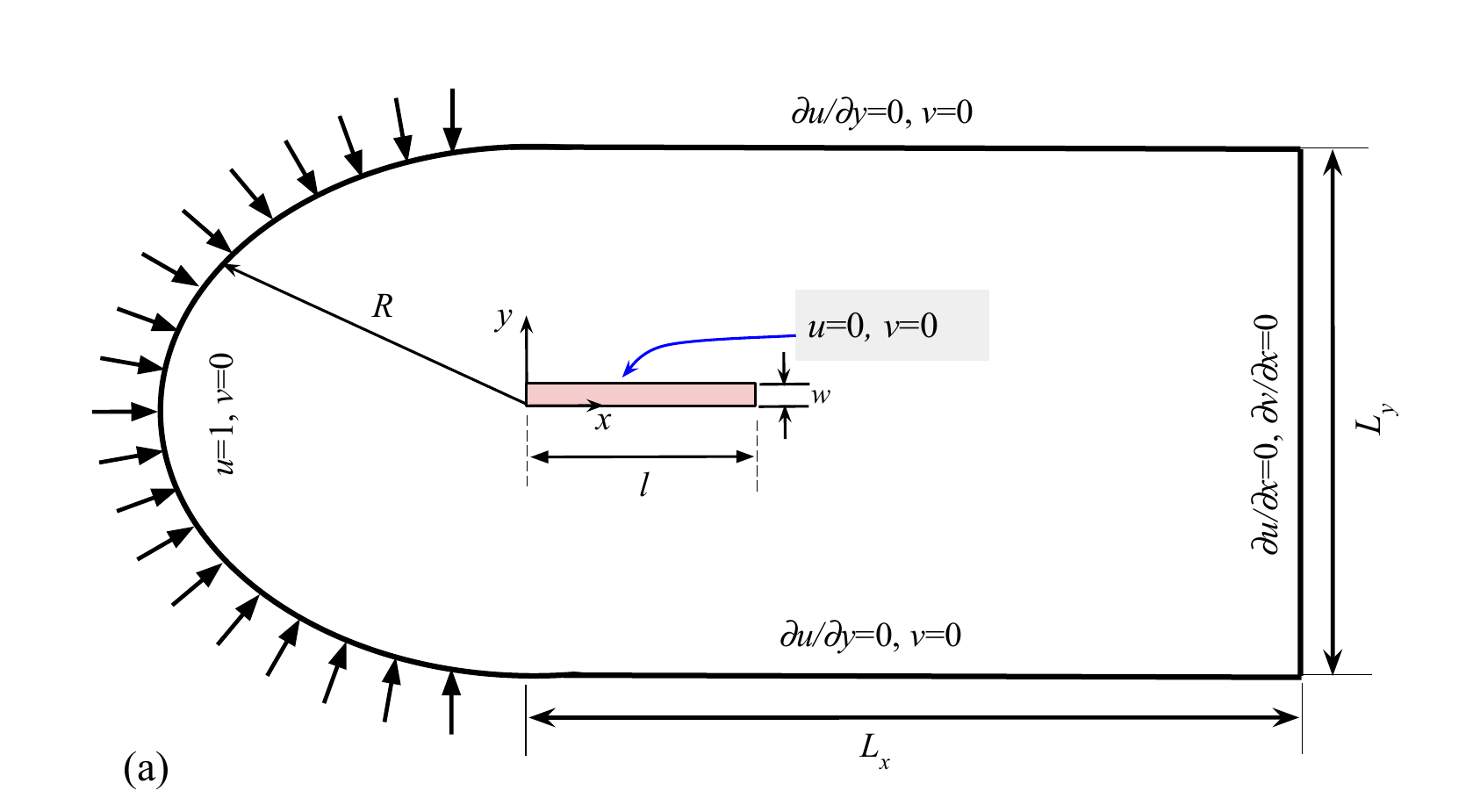}
    \includegraphics[width=1\linewidth]{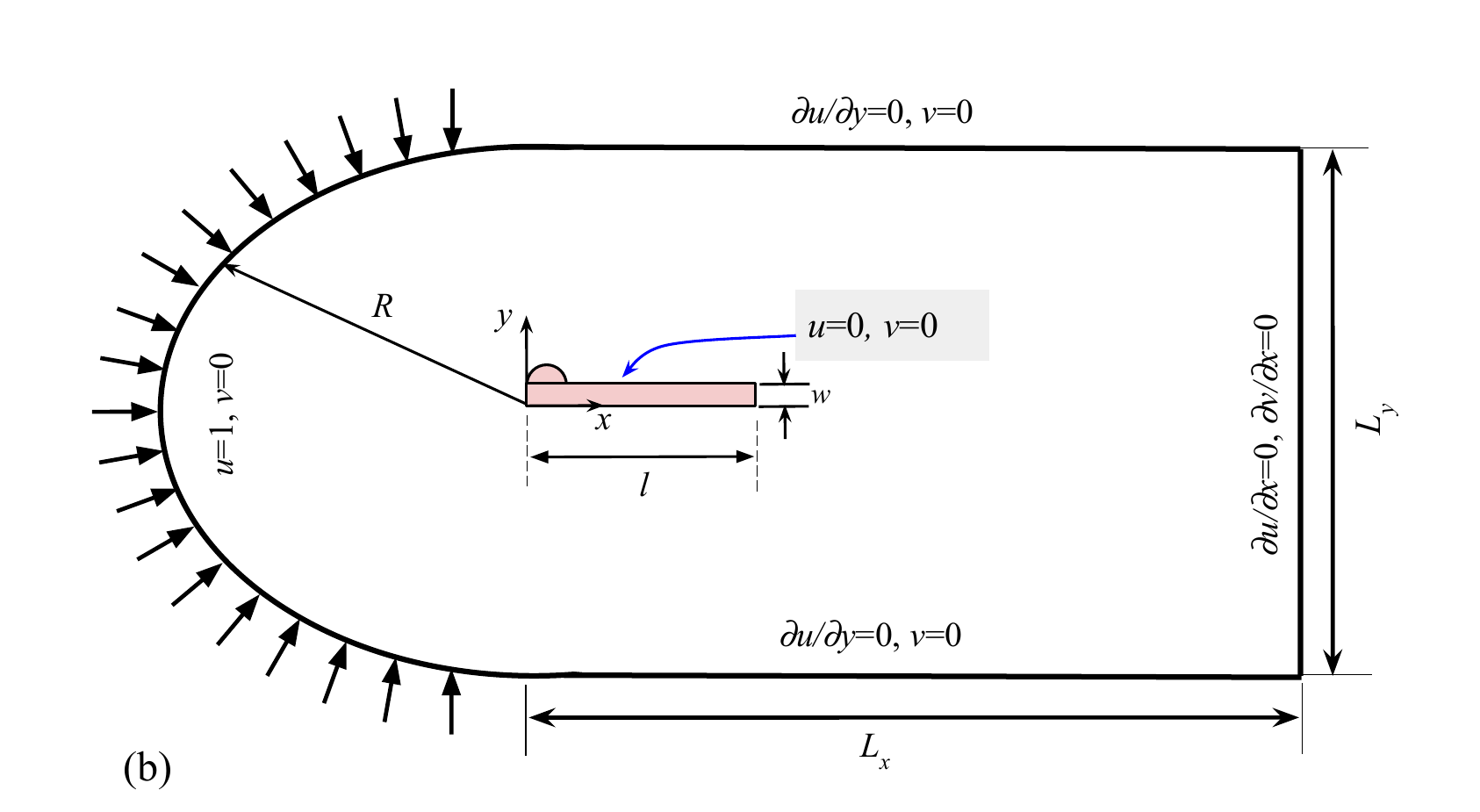}
    \caption{(a) Domain for flat plate. The thickness, \textit{w}, is 5\% of length, \textit{l}. (b) Domain for flat plate with curved leading edge. The thickness, \textit{w}, and the curvature radius is 5\% of length,\textit{l}. }
    \label{fig:schematic}
\end{figure}

This paper examines flow around flat plate with and without curvature (protuberance) at the leading edge at different angle of attach at $Re = 1000$. We compare a plain flat plate with an identical plate carrying a two-dimensional semicircular protuberance of radius $R$ on the upper surface at the leading edge (figure~\ref{fig:schematic}), over a range of angles of attack. At $\alpha = 0$ the protuberance makes the nose locally thicker and rounder. The flow accelerates over it, may separate on its rear side and reattach on the plate, so the boundary layer starts from a disturbed state rather than from the clean Blasius condition. At positive angles of attack the protuberance lies in the suction-side flow near the nose, where it changes the effective leading-edge radius and the local camber. It may fix the separation point at its crest instead of at the sharp edge, and it can make the bubble shorter or longer than on the plain plate. Since the stall type is set by the geometry of this small region, even such a minor change may alter the stall behaviour. Whether the protuberance improves the flow, like a tubercle, or degrades it, like an ice ridge, cannot be decided in advance. A direct comparison with the plain plate under identical conditions is therefore required.

\section{Methodology}
The schematic diagram of the domain with its boundaries is shown in figure 1. The computational domain consists of boundaries and a flat plate of dimension $l\times w$, where $l$ is the length and $w$ is the width of the flat plate. The inlet boundary is kept at a distance of $R$ from the front side of the plate, while the distance  between the outlet boundary and the front side of the plate is $L_x$. The distance from the top/bottom boundary is $L_y$ and freestream velocity is represented as $U_{\infty}$. 

The flow is considered unsteady, incompressible and laminar and it is governed by dimensionless Navier-Stokes equation written as
\begin{equation} \label{eq0:continuity}
\frac{\partial u}{\partial x} + \frac{\partial v}{\partial y} =0
\end{equation}
\begin{equation} \label{eq0:momentum-x}
\frac{\partial u}{\partial t} + u\frac{\partial u}{\partial x} + v\frac{\partial u}{\partial y}= -\frac{\partial P}{\partial x} +\frac{1}{Re}\left(\frac{\partial^2 u}{\partial x^2} + \frac{\partial^2 u}{\partial y^2}\right)
\end{equation}
\begin{equation} \label{eq0:momentum-y}
\frac{\partial v}{\partial t} + u\frac{\partial v}{\partial x} + v\frac{\partial v}{\partial y}= -\frac{\partial P}{\partial y} +\frac{1}{Re}\left(\frac{\partial^2 v}{\partial x^2} + \frac{\partial^2 v}{\partial y^2}\right)
\end{equation}
where \textit{u}, \textit{v},  \textit{P} and \textit{t} denotes the non-dimensional streamwise velocity, transverse velocity, pressure and time, respectively. The non-dimensional variables in eq. (\ref{eq0:continuity}-\ref{eq0:momentum-y}) are defined as $x=x^\star/l$,  $y=y^\star/l$,  $u=u^\star/U_\infty$, $v=v^\star/U_\infty$, $P=P^\star/\rho U_\infty^2$ and $t=t^\star U_\infty/l$. Here $U_\infty$ is the freestream inlet velocity, \textit{l} is the flat plate base length and the superscript $\star$ represent the dimensional quantities. The Reynolds number is defined as Re$=U_\infty l/\nu$, where $\nu$ is the kinematic viscosity.

A finite volume-based solver, Ansys Fluent 2025, has been used to solve the above governing equations (\ref{eq0:continuity}-\ref{eq0:momentum-y}). The second order central differencing schemes is used to discretize the diffusion term, and, the advective terms are discretized using second order upwind schemes. 

\subsection{Domain-independence}

A domain-independence  was performed to ensure that the predicted aerodynamic forces were not influenced by the extent of the computational domain. Three domain sizes were considered by progressively increasing the radial extent, $R$, and the streamwise extent, $L_x$. The three domains were defined by $(R,L_x)=(10,30)$, $(15,40)$, and $(20,50)$, respectively. The drag coefficient at $\alpha=0^{\circ}$ was used to assess the sensitivity of the solution to the domain size. The results for both configurations are summarized in Table~\ref{tab:domain_independence}.

\begin{table}[htbp]
\centering
\caption{Domain-independence study at $\alpha=0^{\circ}$ for the configurations with and without leading-edge curvature.}
\label{tab:domain_independence}
\begin{tabular}{cccc}
\hline
$R$ & $L_x$ & 
$C_D$ without curvature & 
$C_D$ with curvature \\
\hline
10 & 30 & 0.12400 & 0.1332 \\
15 & 40 & 0.1241 & 0.1335 \\
20 & 50 & 0.12397 & 0.1331 \\
\hline
\end{tabular}
\end{table}

For the configuration without leading-edge curvature, the drag coefficient changed from $0.12400$ for the smallest domain to $0.1241$ for the intermediate domain and $0.12397$ for the largest domain. The relative changes between the first and second domains and between the second and third domains were approximately $0.032\%$ and $0.008\%$, respectively. The difference between the smallest and largest domains was approximately $0.024\%$.

Similarly, the configuration with leading-edge curvature showed very little sensitivity to the domain dimensions. The corresponding drag coefficients were $0.1332$, $0.1335$, and $0.1331$. The relative changes between successive domain sizes were approximately $0.075\%$ and $0.150\%$, respectively, while the difference between the smallest and largest domains was approximately $0.075\%$.

The very small variation in $C_D$ with increasing domain size indicates that the computational boundaries have a negligible influence on the predicted aerodynamic forces. The intermediate domain, with $R=15$ and $L_x=40$, was therefore selected for the subsequent simulations.

\subsection{Grid-independence}

A grid-independence was carried out to assess the sensitivity of the numerical results to the spatial resolution. Three successively refined meshes were considered for both geometries. For the plate with the curved leading-edge geometry, meshes M1, M3, and M5 contained 65,621, 82,021, and 110,884 elements, respectively. The corresponding numbers of nodes were 34,146, 42,399, and 56,909. For the plate without the curved leading edge, meshes M2, M4, and M6 contained 67,383, 85,773, and 115,797 elements, with 35,020, 44,272, and 59,369 nodes, respectively. The mesh characteristics are summarized in Table~\ref{tab:grid_independence}.

\begin{table}[htbp]
\centering
\caption{Grid-independence study for the configurations with and without the curved leading edge.}
\label{tab:grid_independence}
\begin{tabular}{cccccc}
\hline
Configuration & Mesh & Nodes & Elements & $C_D$ & $C_L$ \\
\hline
With curvature    & M1 & 34,146 & 65,621  & 0.1330 & 0.132 \\
                  & M3 & 42,399 & 82,021  & 0.1335 & 0.133 \\
                  & M5 & 56,909 & 110,884 & 0.1338 & 0.131 \\
\hline
Without curvature & M2 & 35,020 & 67,383  & 0.1240 & 0.0\\
                  & M4 & 44,272 & 85,773  & 0.1241 & 0.0 \\
                  & M6 & 59,369 & 115,797 & 0.1239 & 0.0 \\
\hline
\end{tabular}
\end{table}

For the configuration with the curved leading edge, the drag coefficient changed from $C_D=0.1330$ on M1 to $0.1335$ on M3 and $0.1338$ on M5. The relative difference between M1 and M3 was approximately $0.38\%$, while that between M3 and M5 was only $0.22\%$. The overall variation in $C_D$ between the coarsest and finest meshes was approximately $0.60\%$. The lift coefficient showed similarly limited sensitivity to the mesh resolution, with values of $0.132$, $0.133$, and $0.131$ for M1, M3, and M5, respectively. The maximum difference between two successive meshes was approximately $1.50\%$.

For the configuration without the curved leading edge, the drag coefficient was $0.1240$, $0.1241$, and $0.1239$ for meshes M2, M4, and M6, respectively. The relative change from M2 to M4 was approximately $0.08\%$, while the change from M4 to M6 was approximately $0.16\%$. The overall variation between the coarsest and finest meshes was therefore less than $0.1\%$.

These results show that further mesh refinement produces only a small change in the predicted aerodynamic force coefficients. In particular, increasing the number of elements from 82,021 to 110,884 for the curved configuration changed $C_D$ by only $0.22\%$, while increasing the number of elements from 85,773 to 115,797 for the configuration without curvature changed $C_D$ by approximately $0.16\%$. The intermediate meshes M3 and M4 were therefore considered sufficiently grid independent and were used for the subsequent simulations.

\subsection{Validation}

The numerical methodology was validated against the two-dimensional
numerical simulation results of \citet{yang2012} for flow
past an inclined flat plate at $Re=1000$. The validation geometry was
chosen to reproduce the configuration considered by \citet{yang2012},
with a thickness-to-chord ratio of $w/l=0.02$, sharp leading and
trailing edges, and an angle of attack of $\alpha=25^{\circ}$. The
simulation was performed at the same Reynolds number using the
transient incompressible Navier--Stokes equations without a turbulence
model.

The present simulation yielded time-averaged force coefficients of

$$
\overline{C_D}=0.5773,
\qquad
\overline{C_L}=1.2643,
$$

compared with the two-dimensional numerical simulation values reported by
\citet{yang2012},

$$
\overline{C_D}=0.5827,
\qquad
\overline{C_L}=1.3756.
$$

The corresponding differences are approximately $0.9\%$ for the drag
coefficient and $8.1\%$ for the lift coefficient. The dominant
wake-velocity frequency obtained from the present simulation was
$f=0.3713~\mathrm{Hz}$. Since $U_{\infty}=1~\mathrm{m\,s^{-1}}$ and
$c=1~\mathrm{m}$, the corresponding Strouhal number is

$$
St=\frac{fc}{U_{\infty}}=0.3713.
$$

This compares with $St=0.3967$ reported by \citet{yang2012}, giving a
difference of approximately $6.4\%$. The close agreement in the mean
drag coefficient, together with the reasonable agreement in the mean
lift coefficient and dominant shedding frequency, indicates that the
present numerical approach captures both the mean aerodynamic loading
and the principal unsteady characteristics of the separated flow.

\section{Results}

\begin{figure}
    \centering
    \includegraphics[width=1\linewidth]{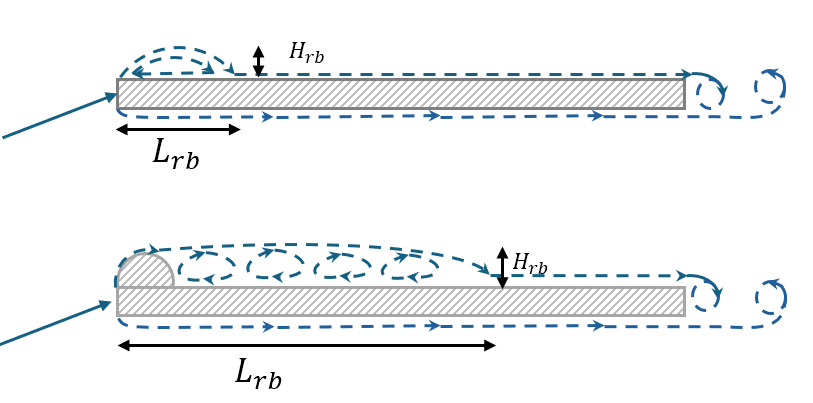}
    \caption{Schematic of the flow with an incident angle of attack  around the flat plate without curvature and with curvature. The separated flow differs in the point of flow separation and the length ($L_{rb}$) and height ($H_{rb}$) of the recirculation bubbles. }
    \label{fig:schema_flow}
\end{figure}

The plates experience the forces due to the flow around them. The differences in the forces acting on the plates with and without leading edge curvature and their temporal variations are direct outcome of the fluid body interaction. The orientation and leading edge geometry are the two important control elements in the present work which influence flow around the plate and affect the forces acting on the plate.  Fig. \ref{fig:schema_flow} shows the schematic of the flow patterns for the flat plate and for the case with curvature at the leading edge. The recirculation bubble's length ($L_{rb}$) and height ($H_{rb}$) are influenced by the curvature and it directly affects the wake characteristics. The following section examines flow profile for one time instant and uses the time varying force coefficients estimated for each scenario of the plate at different angle of attack to investigate the flow physics and associated forces.  

\subsection{Flow field}

\begin{figure*}
    \centering
    \includegraphics[width=0.8\linewidth]{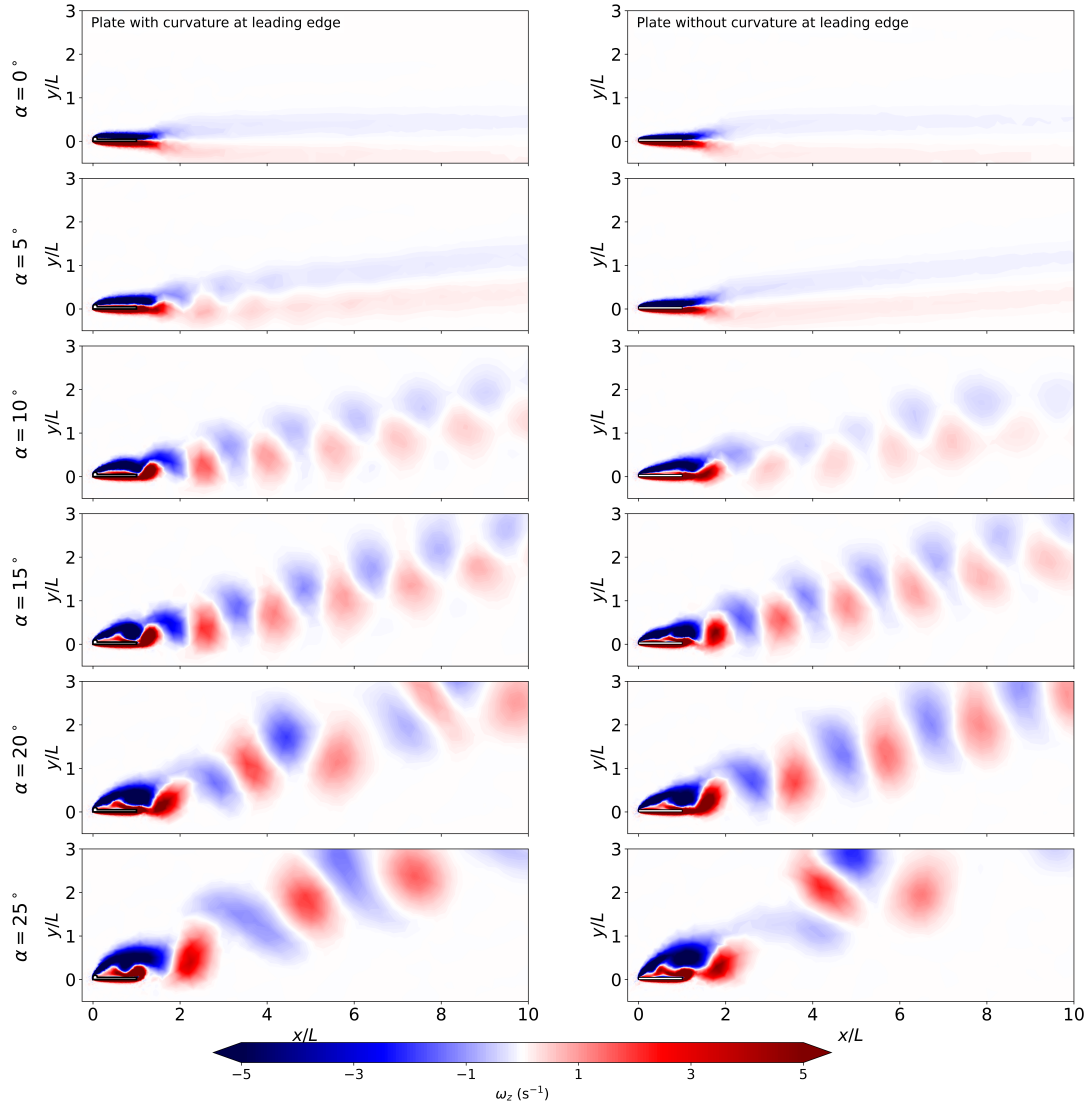}
    \caption{Instantaneous vorticity ($\omega_z$) contours for all angle of attacks for flow around a flat plate and a plate with curvature at the leading edle }
    \label{fig:contours}
\end{figure*}

The flow around the flat plate with and without curvature at the leading edge is influenced by the (i) leading edge and (ii) angle of attack of the flow. Fig. \ref{fig:contours} shows the vorticity contours for flow around the plates.  The flow past flat plates in both the scenarios does not exhibit any shedding at $\alpha = 0^\circ$ ($\alpha$ is the angle of attack). The flat plate with the curvature has a thicker vorticity distribution at the upper face of the plate due to asymmetry in the geometry of the plate. This creates difference in the forces acting on the upper and lower surface of the plate and results in effective lift force. Since the flow distribution before and after the plate is similar for the two scenarios and there is disturbance in the wake, the effective drag force is negligible. When the approaching flow comes at $5^\circ$ angle of attack, the bluffness of the curvature at the leading edge of the plate leads to flow separation with the formation of a detached shear layer which rolls down as vortical structure in the wake. This is the onset of shedding for the flat plate with curvature at the leading edge. These vortices diffuse and only two parallel shear layers are seen beyond an axial distance two times the length of the plate. The plate without curvature does experience a flow separation but the flow get reattached to the sides and no vortical shedding is noticed in the wake. At higher angle of attacks $\alpha \geq 10^\circ$ both the scenarios show alternating von karman vortical shedding.  The deflection of the vortex street is more for the plate with leading edge with a curvature at $\alpha  \in [5^\circ, 10^\circ, 15^\circ, 20^\circ] $.  The flow around the plate with curvature at the leading sedge shows the breaking of vortex street for $\alpha = 20^\circ$.  The disturbance is more pronounced for flow around the plate without curvature at $\alpha = 25^\circ$.  

These wake behaviors are result of the fluid body interaction and they define the nature of the forces acting on the plates over a period of time. The following section examines the time-history of the forces acting on the plates and reports the mean force coefficients. 

\subsection{Force coefficients}

These distinct configurations of leading edge for simple geometry of a flat plate affect the engineering parameters: drag coefficients, lift coefficients and Strouhal Numbers. The forces are influenced by obstacle, shape of the flow separating region and orientation of the flat plate with respect to the flow. The present work for flow around a flat plate at different angle of attack vis-a-vis a flat plate with a curved protrusion at the leading edge reveals the differences in the fluid and obstacle interaction.  The forces acting on the flat plates (both the cases) show different temporal history. The following section discusses the time series for different flow scenarios.  

\subsubsection{Force Coefficient Time series}

\begin{figure}
    \centering
    \includegraphics[width=1\linewidth]{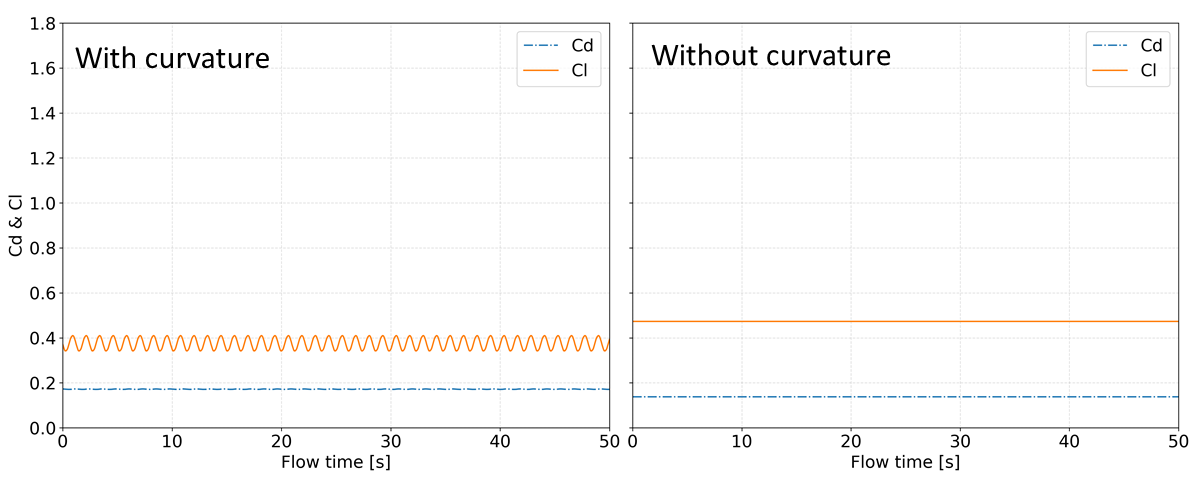}
    \caption{Force Coefficient time series for a flat plate with and without curvature at leading edge placed at $5^\circ$   angle of attack.}
    \label{fig:TS_5degree}
\end{figure}

The force coefficients exhibit steady behavior at $0^\circ $ with noticeable lift coefficient for the flat plate with leading edge curvature due to bluff body effect of the protrusion (curvature) at the leading edge. The case with curvature at the leading edge shows periodic shedding at $AOA = 5^\circ$ whereas the lower angle of attack for the simple case leads to relatively higher $Cl$ and lower $Cd$ while still experiencing a steady effect as shown in Fig. \ref{fig:TS_5degree}. 

\begin{figure}
    \centering
    
    \includegraphics[width=1\linewidth]{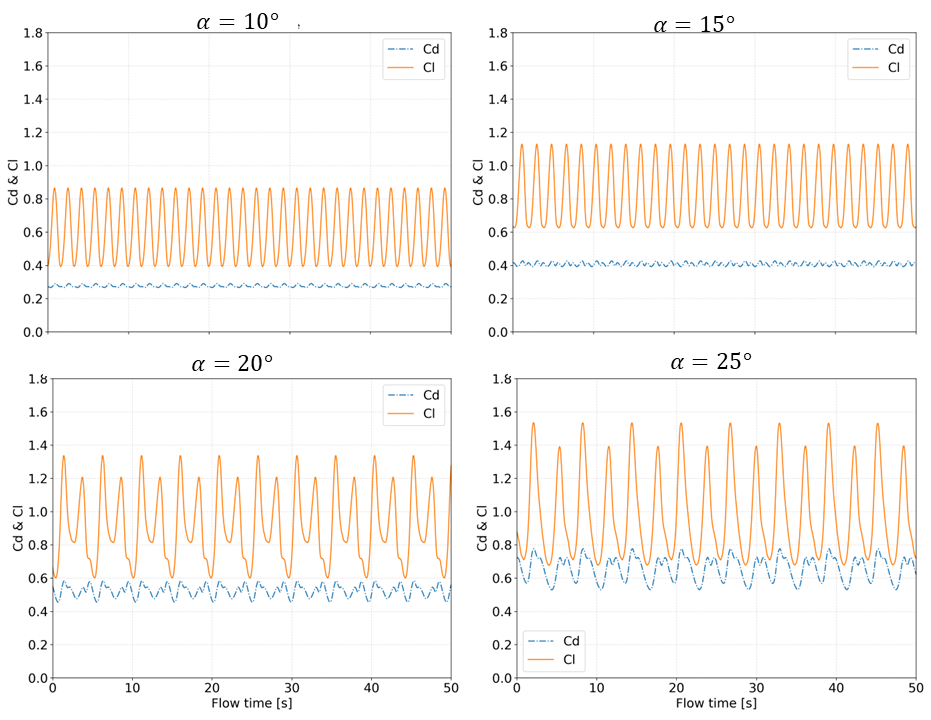}
    \caption{Force Coefficient time series for a flat plate flat with curvature at leading edge placed at different angle of attack}
    \label{fig:TS_WC}
\end{figure}

\begin{figure}
    \centering
    \includegraphics[width=1\linewidth]{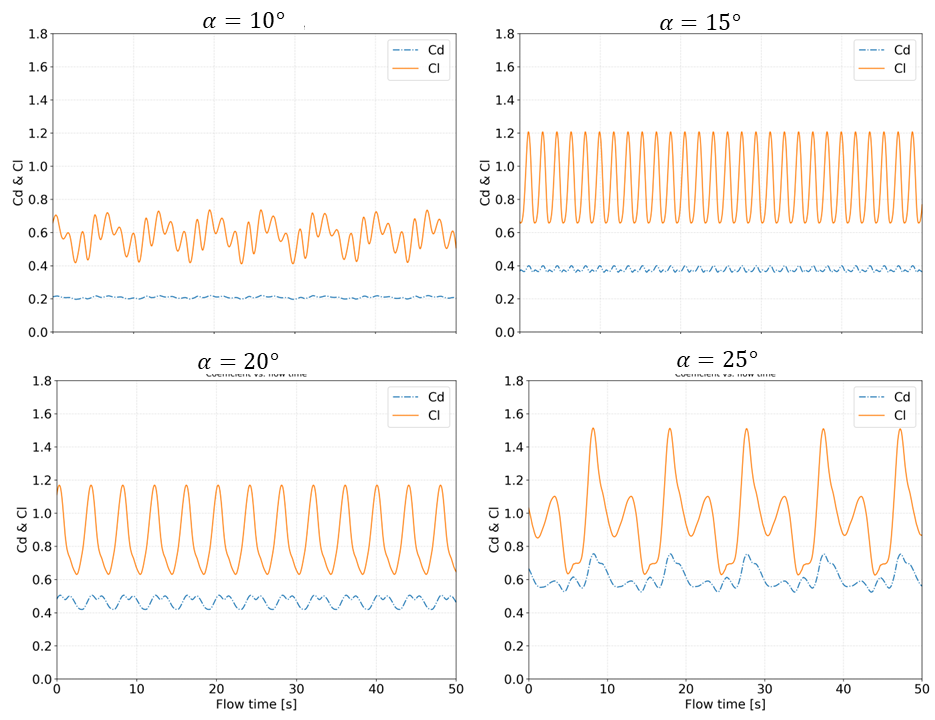}
    \caption{Force Coefficient time series for a flat plate flat without curvature at leading edge placed at different angle of attack}
    \label{fig:TS_WOC}
\end{figure}

Figs. \ref{fig:TS_WC}  and \ref{fig:TS_WOC} show the time series for the force coefficients at various of angle of attack AOA ($ \alpha = 10^\circ$, $15^\circ$, $20^\circ$ and $25^\circ$) for flat plate with a curvature at leading edge and simple flat plate. 
Fig. \ref{fig:TS_WC}  shows an increase in the amplitude for lift coefficient time series when the AOA is changed from $\alpha = 10^\circ$ to $\alpha = 20^\circ$. The range for the lift coefficient is relatively same for the higher AOA ($\alpha = 20^\circ$ and $25^\circ$).  It is noted that the time period increases for lift coefficient as the AOA increases. At higher AOA ($\alpha = 20^\circ$ and $25^\circ$) multiple frequencies is noted during shedding.   For a flat plate with curvature at the leading edge the drag coefficients exhibit steadiness till $\alpha = 5^\circ$, when the AOA is changed to $10^\circ$ periodicity in drag is noted, the amplitude increases and frequency decreases for the drag coefficient time series, when the AOA is increased. At higher AOA the drag coefficient too hint at the high unsteadiness in the shedding. 

The temporal variation of lift coefficient, shows higher unsteadiness with multiple a dominant shedding frequency and many smaller ones at $\alpha = 10^\circ$ .  As the angle of attach is increased to $\alpha = 15^\circ$ and $20^\circ$  periodic shedding is observed for the case with out curvature. The shedding frequency is higher for $\alpha = 15^\circ$. On further increasing the angle of attach ($\alpha=25^\circ$), multiple frequencies are noted along with large variation in the lift coefficients over a period.  The drag coefficients show unsteady trend for $\alpha \geq 15^\circ$ due to fluctuations in pressure, downstream of the sharp leading edge of flat plate without any curvature.  The drag coefficients are lower for flat plate without leading edge curvature (Fig. \ref{fig:TS_WOC}) as to curved bluff body shaped leading edge (Fig. \ref{fig:TS_WC}). 

The differences in force coefficient's temporal history owes to the flow separation region at the leading edge. Larger drag coefficient ranges were noted for the flat with curvature at the leading edge due to the bluffness at the edge.  The temporal profiles show early onset of periodic shedding for the case with curvature.  The plate without the curvature shows a multiple frequency shedding at $\alpha = 10^\circ$ . The effects can be further investigated when we examine the combined temporal profile of both the drag and lift coefficients using $Cl$-$Cd$ phase plots. 

\subsubsection{Force Coefficient Phase plots}

\begin{figure}
    \centering
    \includegraphics[width=1\linewidth]{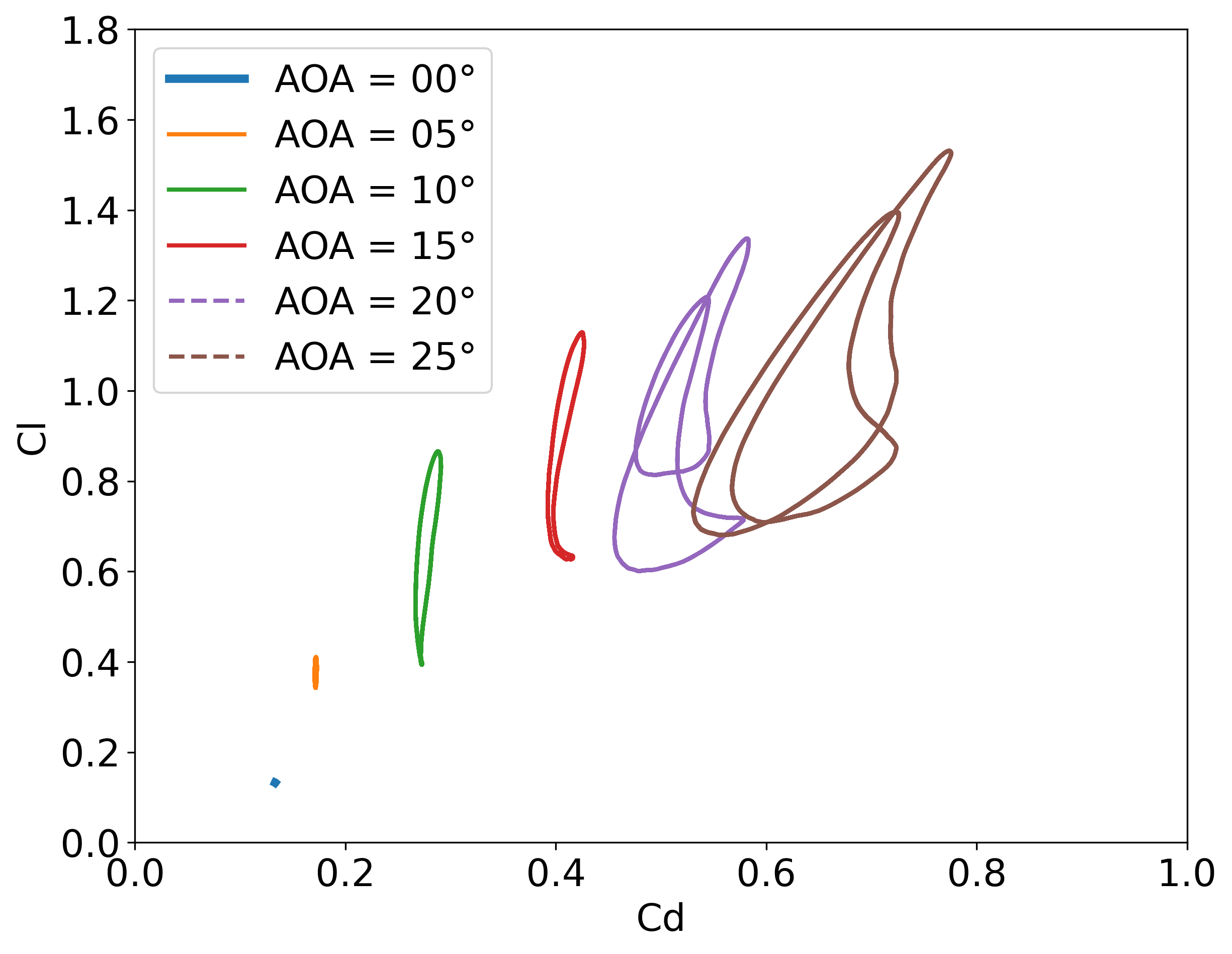}
    \caption{Cd-Cl Phase plot for flat plate with curvature at the leading edge  placed at different angle of attack}
    \label{fig:CDCL_WC}
\end{figure}

\begin{figure}
    \centering
    \includegraphics[width=1\linewidth]{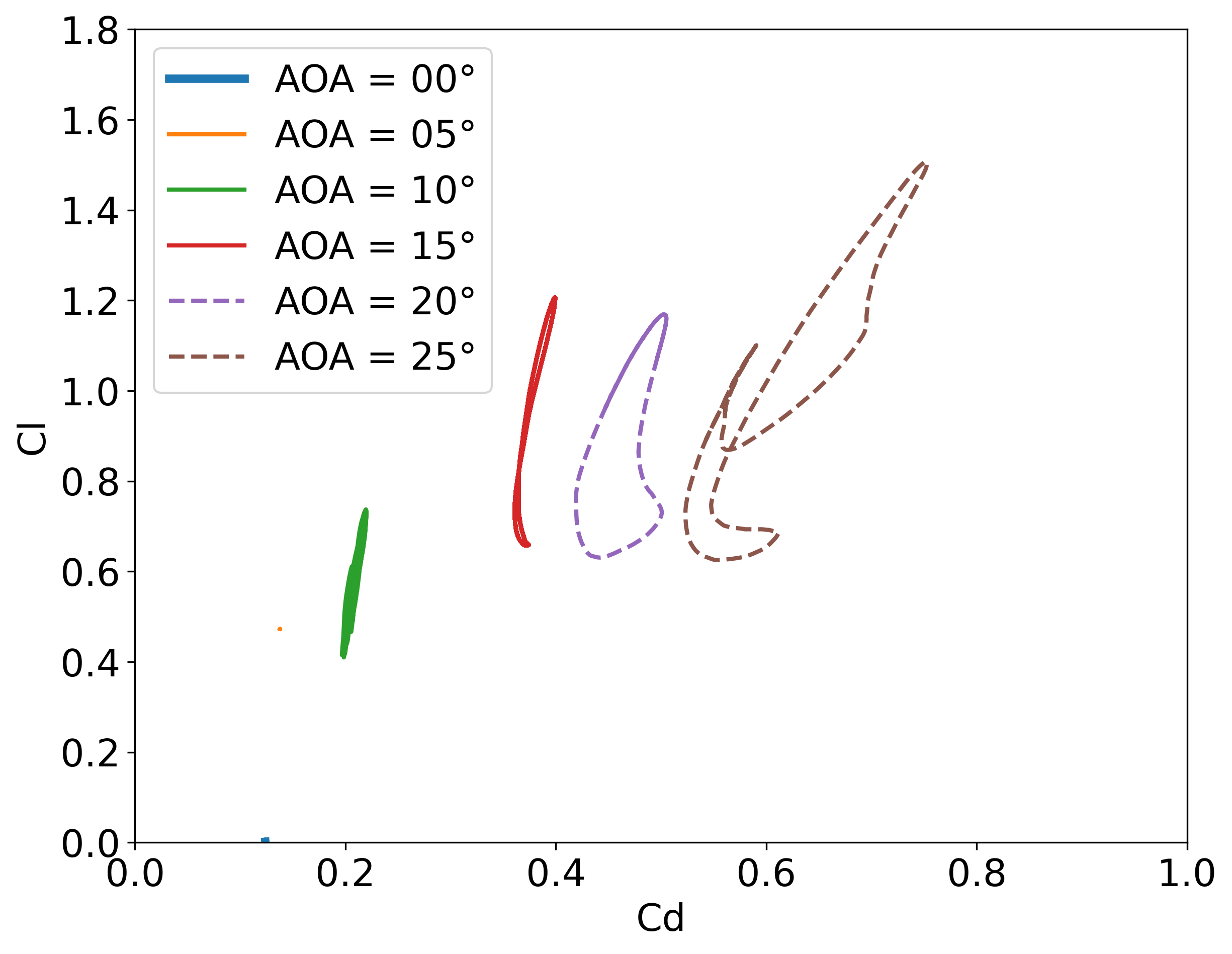}
    \caption{Cd-Cl Phase plot for flat plate flat without curvature at leading edge with curvature placed at different angle of attack}
    \label{fig:CDCL_WOC}
\end{figure}

The coupled effect of the drag and lift forces on the flat plate scenarios can be explained using the $Cl-Cd$ phase plots. These phase plots help establish the inter-play of the forces along with their time varying behavior.  Fig. \ref{fig:CDCL_WC} shows the $Cl-Cd$  phase plots for the flat plate case with curvature at the leading edge. It is observed that the $Cl-Cd$ plot for AOA $\alpha = 0^\circ$ is a point which hints at the steady flow behavior of the fluid. As the AOA is increased to $\alpha = 5^\circ$, the $Cd$ remains unchanged but the onset of shedding is evident from the spread of $Cl$. At larger angle of attacks it is noted that a closed loop is covered by the $Cl-Cd$ curve indicating the periodicity in the flow over the plate with curvature at the leading edge. The single loop of $Cl-Cd$ curve transitions to two loops for higher AOA ($\alpha = 20^\circ$ and $25^\circ$). The $Cl-Cd$ dependence is different for the case with a flat plate without any curvature at the leading edge (Fig. \ref{fig:CDCL_WOC}). The flow remains steady and non-shedding till $\alpha = 5^\circ$ and shows shedding when the AOA is increased to $\alpha = 10^\circ$. A single loop is observed till $\alpha = 20^\circ$. The $C;-Cd$ curve shows two loops for the higher AOA of $25^\circ$. The different trends in the phase plots hint at the varying dynamical nature of the force coefficients at different angle of attacks.  A Poincare map was generated for each shedding case to explain the nature of shedding.

The Poincare map was generated using the $Cl-Cd$ dynamics especially focusing on the upward crossing times ($t_1,t_2,t_3,\ldots$) for $Cl$ in the $Cl$ time series and noting the corresponding $Cd$ values as section information.  The idea was to use $Cl$ to trigger sampling and $Cd$ to measure the state at the sampling instant.  The sampling condition uses $Cl$ and record the $Cd$ values at the sampling instants. 
$$ 
Cl(t_n)=\overline{Cl},  
\qquad  
\frac{dCl}{dt}\bigg|_{t_n}>0.  
$$

The recorded drag coefficient for generating the map is
$$ 
Cd_{n}^{\text{section}}=Cd(t_n).  
$$

\begin{figure}
    \centering
    \includegraphics[width=1\linewidth]{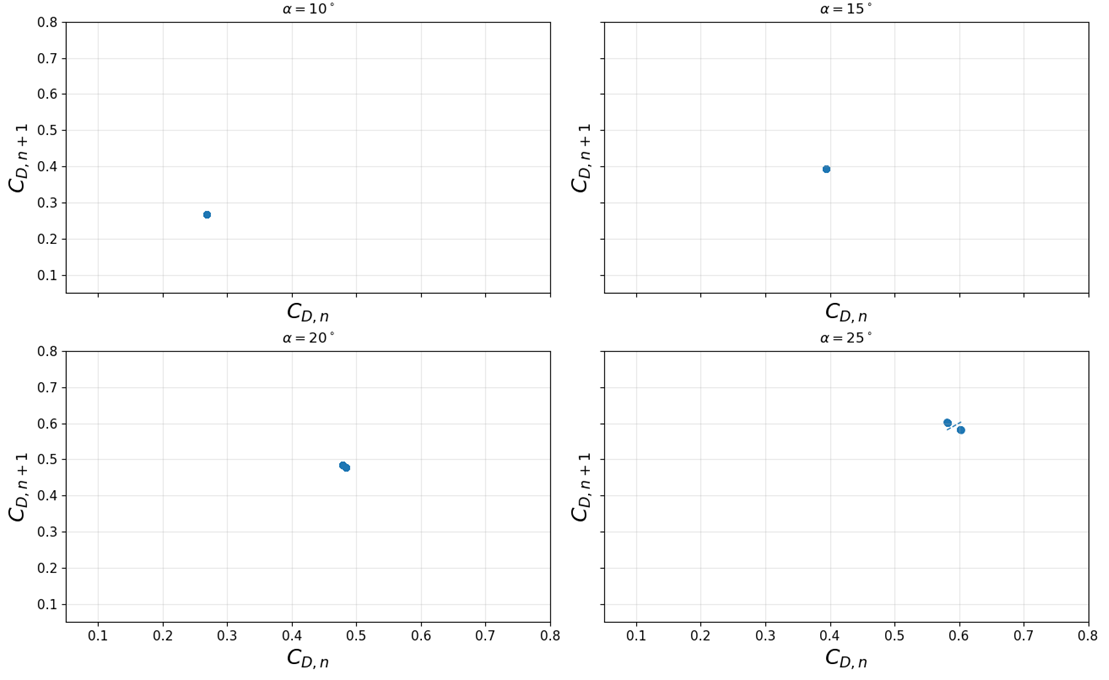}
    \caption{Poincare sets for flat plate flat with curvature at leading edge placed at different angle of attack}
    \label{fig:P_WC}
\end{figure}

\begin{figure}
    \centering
    \includegraphics[width=1\linewidth]{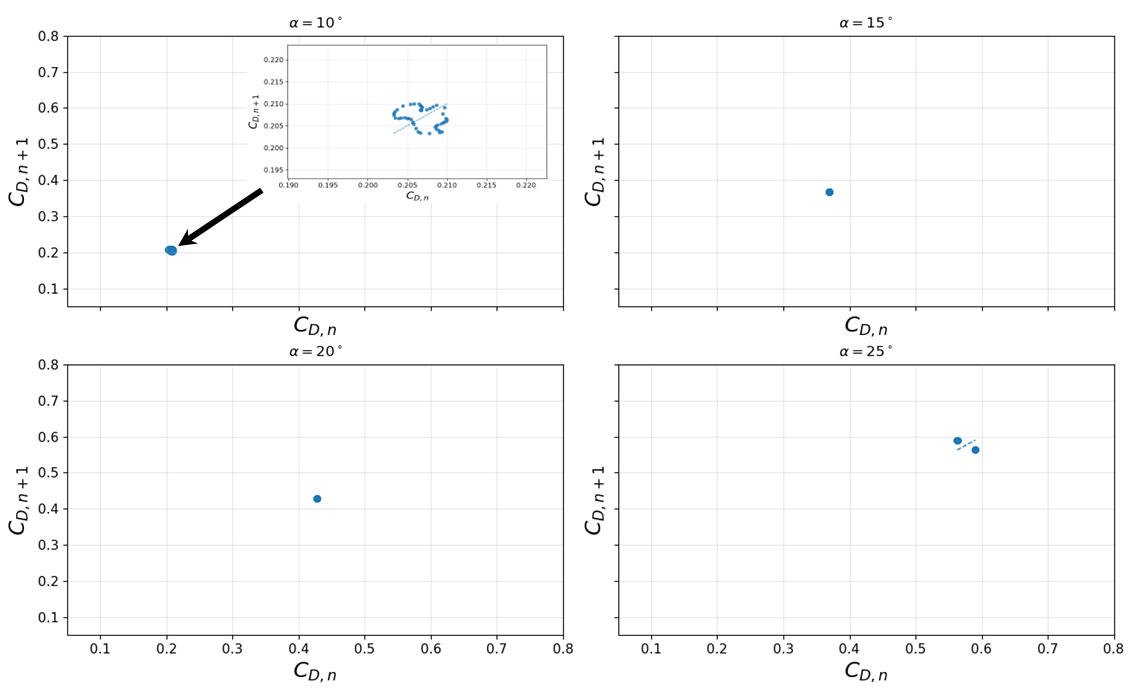}
    \caption{Poincare sets for flat plate flat without curvature at leading edge placed at different angle of attack}
    \label{fig:P_WOC}
\end{figure}

The returned map plots (Figs. \ref{fig:P_WC} and \ref{fig:P_WOC})
$$  
Cd_{n+1}^{\text{section}}  
\quad \text{versus} \quad  
Cd_{n}^{\text{section}}.  
$$ 
Fig. \ref{fig:P_WC} enables a deeper look into the $Cl-Cd$ phase plots of Fig. \ref{fig:CDCL_WC}. The single points on the Poincare map for $\alpha = 10^\circ$ and $15^\circ$ show the enable periodic limit cycle also reflected as single loop in  Fig. \ref{fig:CDCL_WC}. Two staggered dots on the map for $\alpha = 20^\circ$ hint at period increasing number of frequencies involved. Two distinct $Cd$ values on the Poincare map (Fig. \ref{fig:P_WC} show multi-periodic response of the system at higher AOA.  

The dynamics for flat plate without curvature at the leading edge at an angle of attack $\alpha = 10^\circ$ shows a quasi-periodic behavior when the periodic seeming dot is zoomed Fig. \ref{fig:P_WOC}. This is the signature of the periodic shedding with small frequencies superimposed in the time series (as shown in Fig. \ref{fig:TS_WOC}) for $\alpha = 10^\circ$.  Stable single frequency periodic cycles are noted for $\alpha = 15^\circ$ and $20^\circ$. The multi-period behavior is noted for angle of attack of $\alpha = 25^\circ$. 

\begin{figure}
    \centering
    \includegraphics[width=1\linewidth]{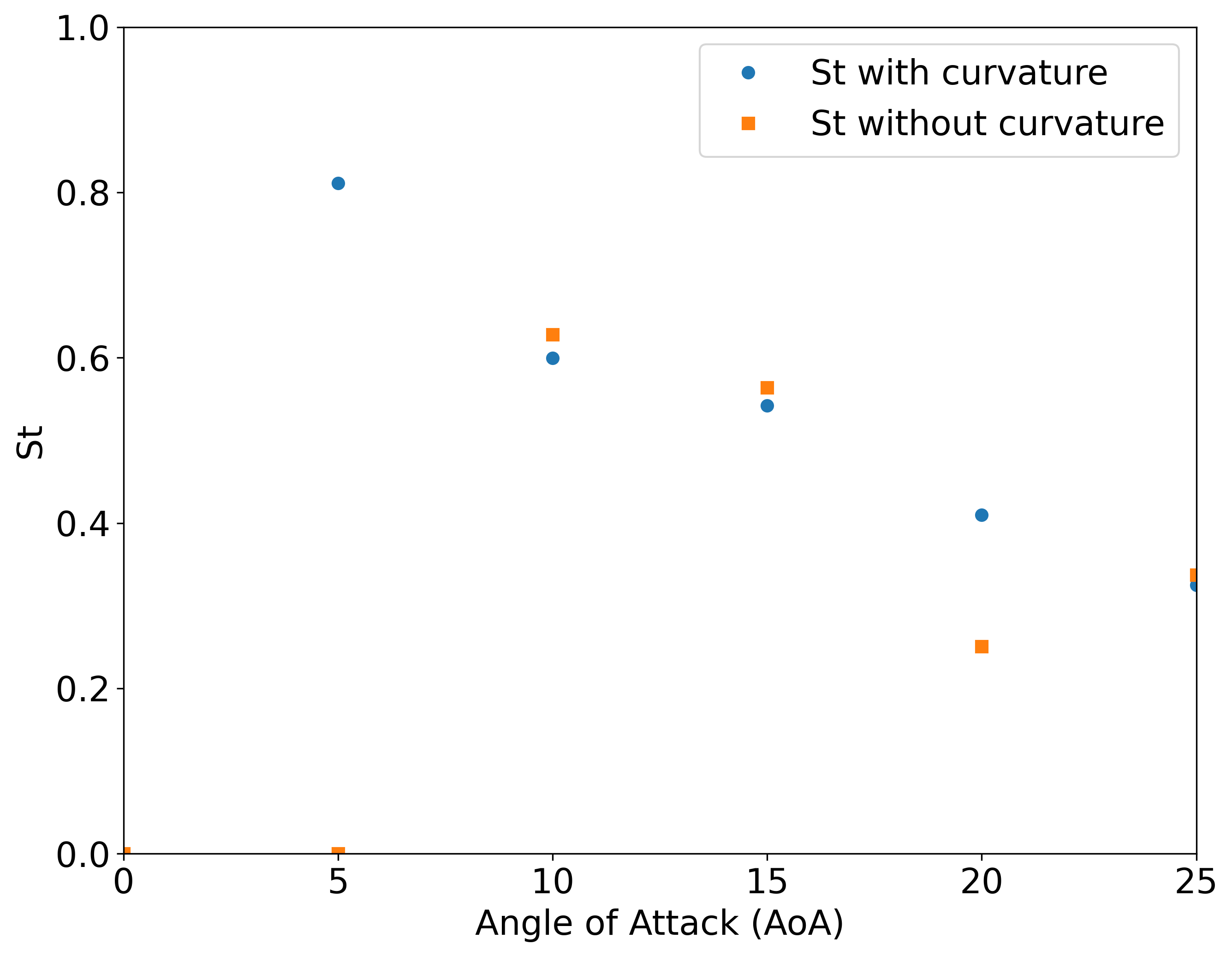}
    \caption{Strouhal Number for flat plate flat with and without curvature at leading edge placed at different angle of attack }
    \label{fig:st}
\end{figure}

The flow dynamics around the flat plates with and without curvature has so far been explained using the temporal variation of the force coefficients. The time series enabled us to examine the nature of shedding and informed us about the number of frequencies involved in the fluid and plate interaction. In order to quantify the shedding Strouhal numbers were estimated using the time series for the lift coefficients. Fig. \ref{fig:st} shows the gradual decrease in the Strouhal numbers for both the leading edge variation (till AOA $\alpha = 20^\circ$).  
The flat plate with curvature at the leading edge experienced shedding at AOA $\alpha = 5^\circ$ whereas the flat plate without curvature had steady non-shedding flow for $\alpha = 0^\circ$ and $5^\circ$.  For angle of attacks $\alpha = 10^\circ$ and $15^\circ$ the shedding frequency for flat plate without curvature is relatively higher and then it dips down for $\alpha = 20^\circ$ due to flow separation and higher mixing near the body. The Strouhal number at $25^\circ$ angle of attack is same for both the cases. It is observed that the shedding frequency has decreases linearly for the flat plate with a curvature at the leading edge but the for the case without curvature there a dip in Strouhal number is noted at $\alpha = 20^\circ$.  

\subsubsection{Statistical Moments for Cl and Cd}

\begin{figure}
    \centering
    \includegraphics[width=1\linewidth]{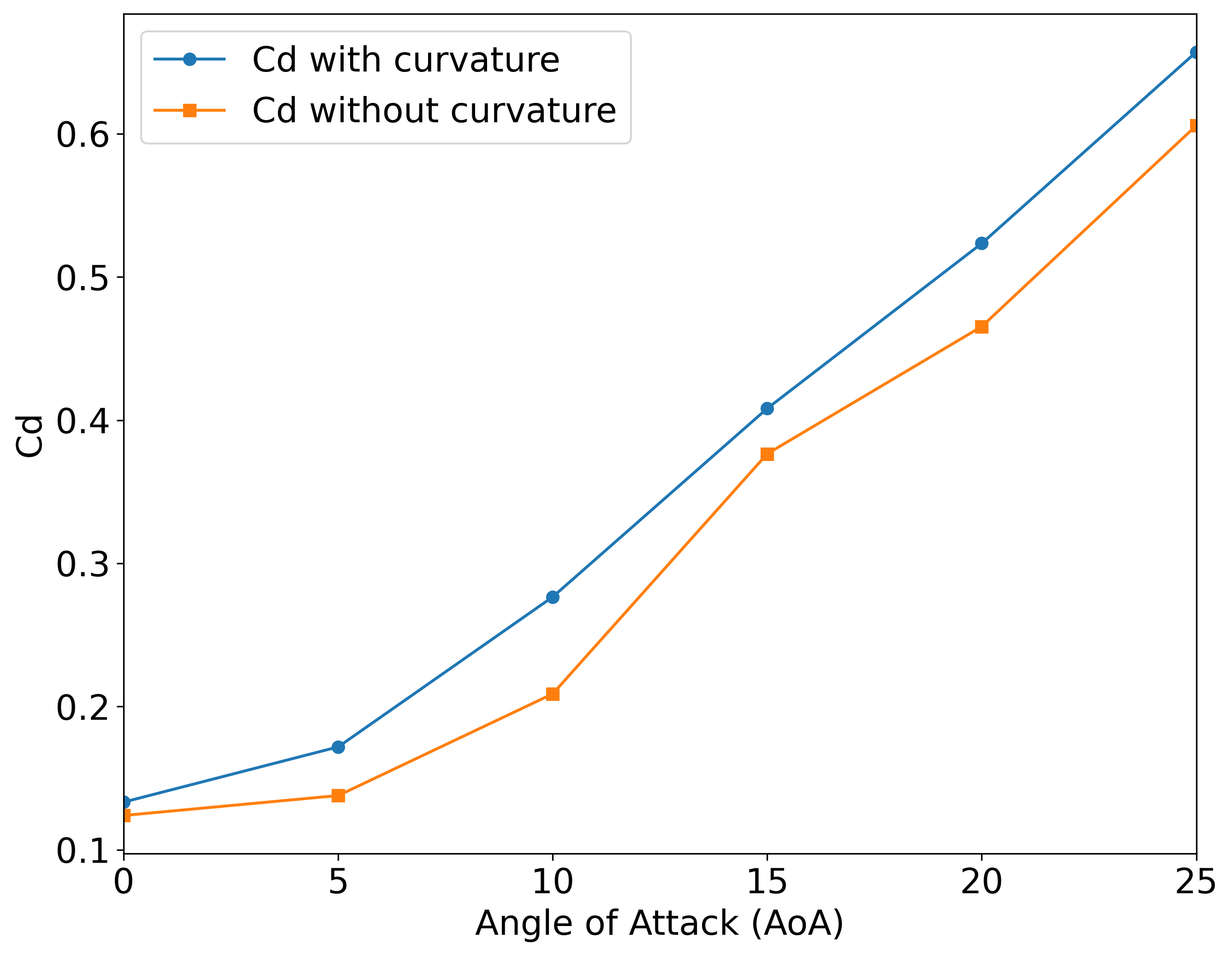}
    \caption{Mean drag coefficient at various angle of attack (AoA)}
    \label{fig:Mean_Cd}
\end{figure}

\begin{figure}
    \centering
    \includegraphics[width=1\linewidth]{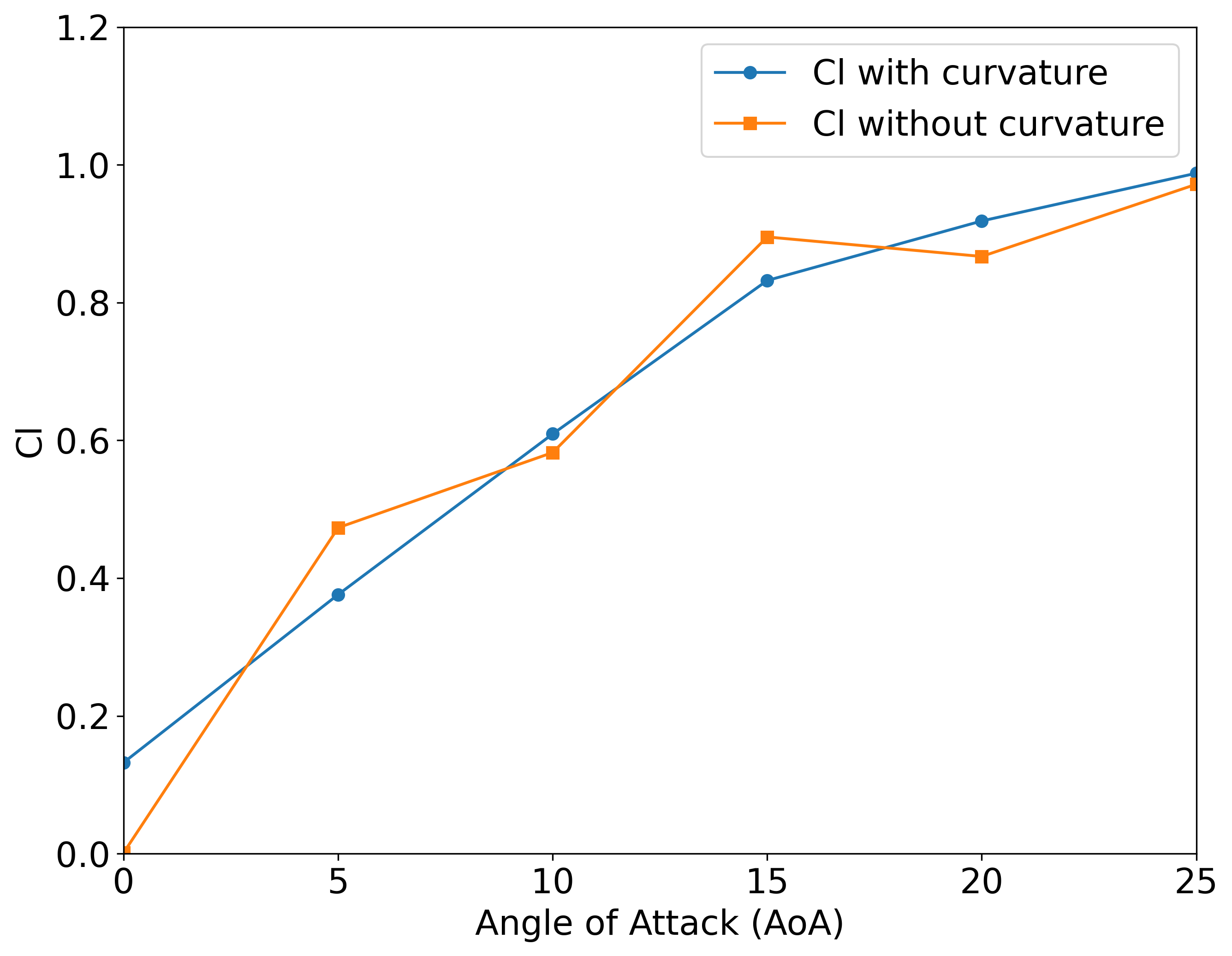}
    \caption{Mean lift coefficient at various angle of attack (AoA)}
    \label{fig:Mean_Cl}
\end{figure}

\begin{figure}
    \centering
    \includegraphics[width=1\linewidth]{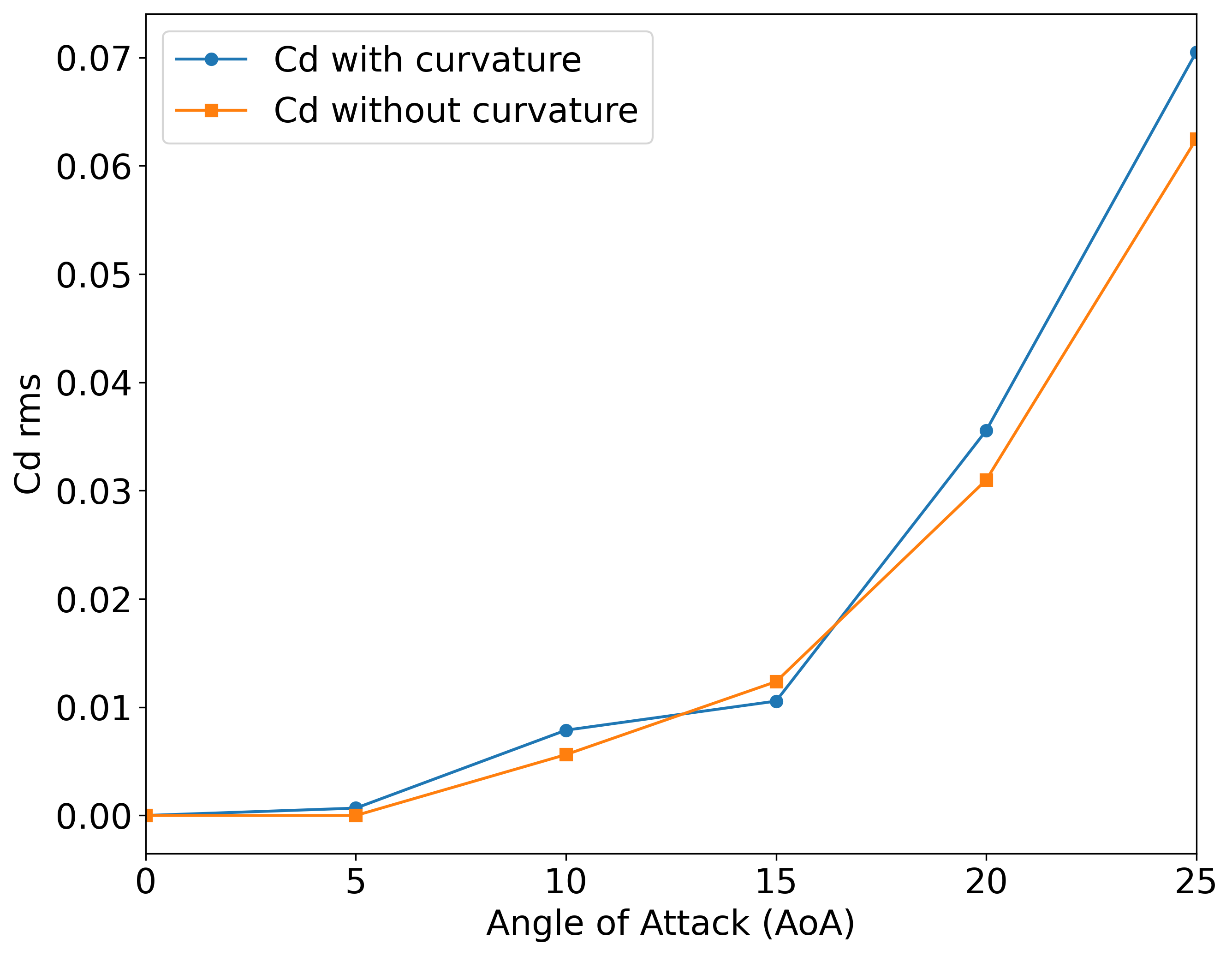}
    \caption{RMS for Cd at various angle of attack (AoA)}
    \label{fig:RMS_Cd}
\end{figure}

\begin{figure}
    \centering
    \includegraphics[width=1\linewidth]{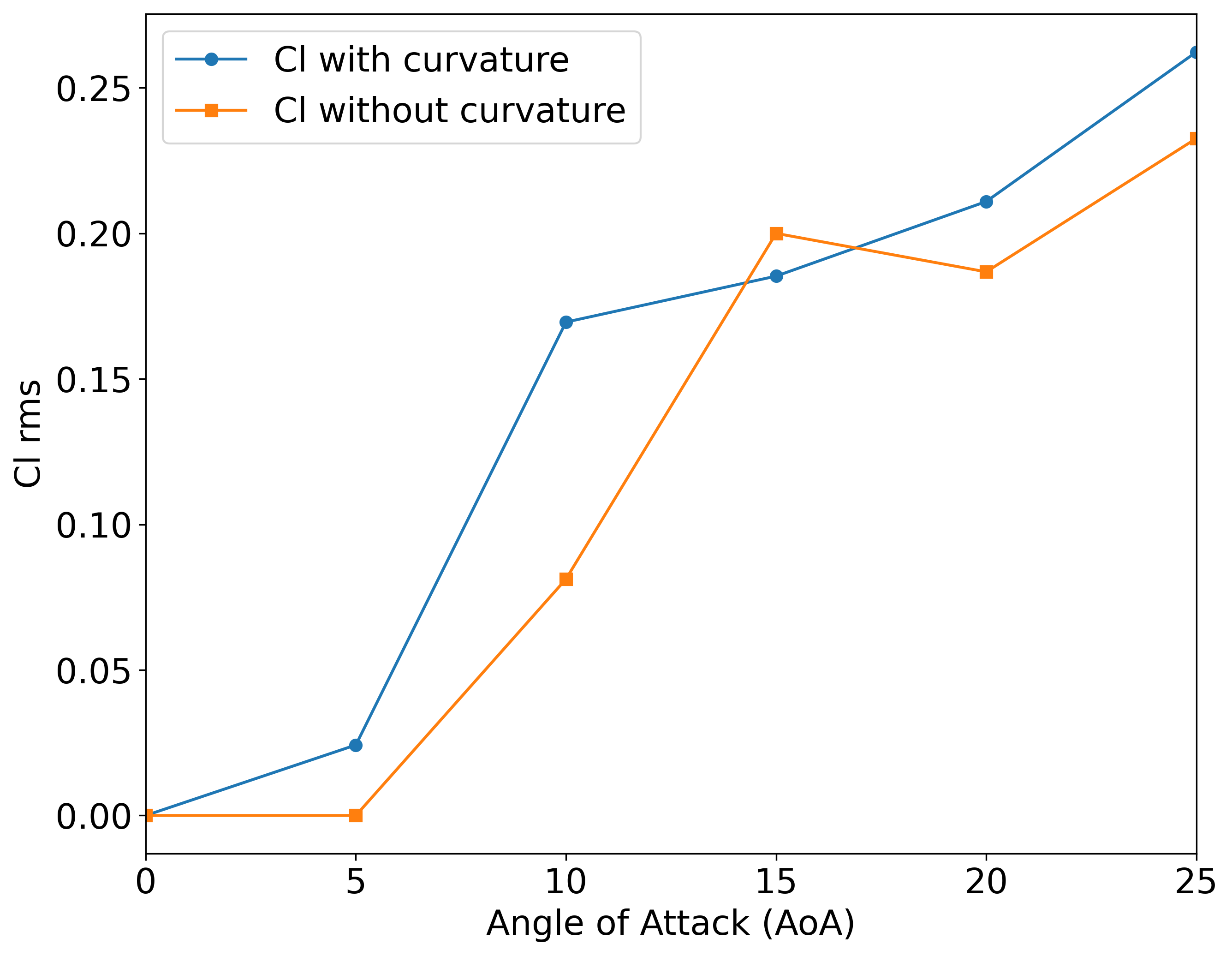}
    \caption{RMS for Cl at various angle of attack (AoA)}
    \label{fig:RMS_Cl}
\end{figure}

A statistical analysis of the time series obtained for the lift and drag coefficients enables insignt into the symmetric nature of the force coefficients along with the strength, asymmetry and intermittency of the  fluctuations involved when the shedding occurs during a flow. The mean values were estimated using $\overline{C} = \frac{1}{N}
\sum_{i=1}^{N} C_i$ where $\overline C$  is the mean value ($Cd$ or $Cl$) and $N$ is the number of samples in the time series.  Fig. \ref{fig:Mean_Cd} shows the mean drag coefficients for the two cases for different angle of attacks.  The leading edge curvature at $0^\circ$ angle of attack provides bluffness at the leading edge as compared to the flat plate without any curvature. Since the plate is parallel to the flow direction the most of the drag contribution is due to the shear stresses at the surface. The flat plate with curvature at the leading edge has an additional form drag contribution due to the bluff end.  As the angle of attack increases the bluffness of the body is combined effect of the shape along with orientation to the flow. A $25^\circ$ angle of attack being the case with maximum bluffness of the body due to orientation of the body.  the increase in drag is directly related to the increased angle of attack for $\alpha \geq 5^\circ$.  
Fig. \ref{fig:Mean_Cl} shows the mean lift coefficients for the two cases for different angle of attacks. At $0^\circ$ angle of attack the flat plate without the curvature has zero lift due to the inherent symmetry of the plate.  The flat plate with curvature at the leading edge shows a lift coefficient $Cl = 0.133$ for $0^\circ$ angle of attack due to the asymmetric geometry of the plate in the flow.  As the angle of attack increases, an increase in the lift coefficient is observed for both the cases. The increase is linear for the flat plate with curvature at the leading edge till $\alpha = 15^\circ$ and then the rate of increase decreases for $\alpha = 20^\circ$ and $25^\circ$. The flat plate without curvature has steady flow for $\alpha = 5^\circ$ with a $Cl = 0.473$ as compared to $Cl = 0.377$ for the case with curvature. The shedding regimes show an increasing trend in the lift coefficients but the values are either greater or lesser than that for the flat plate with curvature. 
 These mean values of the force coefficients provide a glimpse of the average forces acting on the plates but a quantification of strength of fluctuations ($C_i' = C_i-\overline{C}$) will help examine the unsteadiness in the flow due to the plate orientation.  Figs. \ref{fig:RMS_Cd} and \ref{fig:RMS_Cl} show the rms of fluctuations $C_{\mathrm{rms}} = \sqrt{\frac{1}{N}
\sum_{i=1}^{N}
\left(C_i-\overline{C}\right)^2
}$ in the drag and lift force coefficients at different AOA for the flat plates with and without curvature at the leading edge.  It is observed that the both the drag and lift loading on the plates increases with angle of attack.  
Relatively similar drag loading is experienced by both the types of the plates as the angle of attack is increases. Coefficient of lift shows a larger rms at lower AOAs for the plate with curvatures at the leading edge owing to the increased circulation around the plates. For the plate without curvature the  increase is more pertinent for lift coefficients where unsteadiness is noted at $\alpha = 15^\circ$.  The rms decreases for $\alpha  = 20^\circ$ . Similar behavior is noted at $\alpha = 10^\circ$ for the plate with curvature. These angle of attacks are the onset of unsteadiness in the flow.  The intensity of Cl fluctuations reaches $20\%$ or more at higher AOA whereas the $Cd$ fluctuations have a rms value less than $10\%$ at the extreme orientations. 
 
The time series shown in Fig. \ref{fig:TS_WC} and \ref{fig:TS_WOC} show the shedding regimes with different $Cd$ and $Cl$ loadings with time. Estimate of the skewness $Sk_C
=
{(
\displaystyle
\frac{1}{N}
\sum_{i=1}^{N}
\left(C_i-\overline{C}\right)^3
)}/{(
\left[
\displaystyle
\frac{1}{N}
\sum_{i=1}^{N}
\left(C_i-\overline{C}\right)^2
\right]^{3/2}
)}$ of the time series was employed to get a complete picture of the distribution of the fluctuations about the mean values for both the lift and drag coefficients. Table \ref{tab:Cd_summary} shows the skewness for both the plate scenarios. The flat plate case with curvature at the leading edge has maximum positive asymmetry at $\alpha = 10^\circ$ with $Sk_{Cd} = 0.552$ and then it decreases finally show negative skew at $\alpha = 25^\circ$. Large positive skew in the drag coefficient fluctuations are noted at $\alpha = 15^\circ$ and $\alpha = 25^\circ$ for the plate without curvature when $Sk_{Cd} = 0.670$ and $Sk_{Cd} = 0.981$.  The lift coefficient has an increasing positive skew for both the cases (Table \ref{tab:Cl_summary}). The plate with curvature started shedding with nearly symmetric fluctuations at $\alpha = 5^\circ$ and then the positive skew increases to $Sk_{Cl} = 0.625$. The plate without curvature has onset of shedding at $\alpha = 10^\circ$ as shown earlier. The negative skewness is $Sk_{Cl} = -0.224$ at $\alpha = 10^\circ$ and then has positive skewness at higher AOA. The values of skewness for the $C_l$ are indicative of the vortical actions in the wake and explains the increasing lift loading on the plate at different angle of attacks. 

Since the force coefficients' time series have periodic patterns involved, detailed information can be extracted if kurtosis is estimated for the fluctuations. Pearson kurtosis ($Ku(C)
=
{(
\displaystyle
\frac{1}{N}
\sum_{i=1}^{N}
\left(C_i-\overline{C}\right)^4
)}/{(
\left[
\displaystyle
\frac{1}{N}
\sum_{i=1}^{N}
\left(C_i-\overline{C}\right)^2
\right]^2
)}$) was estimated for both the drag and lift coefficients time series for the two flat plate cases investigated in this paper.  The kurtosis values for pure Gaussian distribution is  $Ku_{\mathrm{Gaussian}} = 3$ and for sinusoidal wave is $Ku_{sine} = 1.5$. These limits inform us about the excursions and distributions of the estimated force coefficient values ($Cd$ and $Cl$). Tables \ref{tab:Cd_summary} and \ref{tab:Cl_summary} show the kurtosis values for drag and lift coefficients for the plate with and without curvature at the leading edge at different angle of attack.  Since all the estimates of kurtosis are less than 3, the force coefficients have flatter distribution around the mean and there are no extreme excursions.  Table \ref{tab:Cd_summary} shows the estimated values of $Ku_{Cd}$ between 1.5 to 2.026 with the maximum at $\alpha = 20^\circ$ for flat plate with curvature at the leading edge. The values are relatively higher for flat plate case without curvature.  Except for $\alpha = 20^\circ$ the kurtosis values hint at no uniform wave form at all the other AOAs which is corroborated from the time series of Fig. \ref{fig:TS_WOC} . The time series for Cd is either sinuous or has uniformly distributed amplitudes for late with curvature, this is a signature of vortex shedding from curved surface.  A notable exception is $\alpha = 20^\circ$ which does not show any unique waveform. Even for the lift coefficients (Table \ref{tab:Cl_summary}) for the plate with curvature the kurtosis values are around 1.5 for $\alpha = 5^\circ, 10^\circ$ and $15^\circ$. The values shift to $Ku_{Cl} \equiv 2$ for $\alpha = 20^\circ $ and $25^\circ$ reflects  a wave form different from sinusoids (Fig. \ref{fig:TS_WC}).  The flat plate without curvature has relatively periodic waveform for $\alpha = 15^\circ$ and $20^\circ$ with $Ku_{Cl} = 1.511$ and $1.608$ respectively. The values for the onset of shedding at $\alpha = 10^\circ$ is 2.282 and for the extreme orientation of $\alpha  = 25^\circ$ is 2.686 (closer to Gaussian Ku = 3). 

\begin{table*}[!htbp]
    \centering
\caption{Summary of drag coefficient statistics for the flat plate with and without leading edge}
\label{tab:Cd_summary}
    \begin{tabular}{|c|c|c|c|c|c|c|c|c|}\hline
         &  \multicolumn{4}{|c|}{Plate with curvature at leading edge}&  \multicolumn{4}{|c|}{Plate without curvature at leading edge}\\\hline
         $\alpha$&  $Cd_{mean}$&  $Cd_{rms}$&  $Sk_{Cd}$&  $Ku_{Cd}$&  $Cd_{mean}$&  $Cd_{rms}$&  $Sk_{Cd}$& $Ku_{Cd}$\\\hline
         $0^\circ$&  0.133&  0.000&  -&  -&  0.124&  0.000&  -& -\\\hline
         $5^\circ$&  0.172&  0.001&  0.129&  1.507
&  0.138&  0.000&  -& -\\\hline
         $10^\circ$&  0.277&  0.008&  0.552&  1.879
&  0.209&  0.006&  -0.096& 2.203
\\\hline
         $15^\circ$&  0.408&  0.011&  0.222&  1.874
&  0.376&  0.012&  0.670& 2.022
\\\hline
         $20^\circ$&  0.523&  0.036&  -0.178&  2.026
&  0.465&  0.031&  -0.275& 1.485
\\\hline
         $25^\circ$&  0.657&  0.071&  -0.270&  1.790&  0.606&  0.062&  0.981& 2.752\\ \hline
    \end{tabular}

\end{table*}

\begin{table*}[!htbp]
    \centering
\caption{Summary of lift coefficient statistics for the flat plate with and without leading edge}
\label{tab:Cl_summary}
    \begin{tabular}{|c|c|c|c|c|c|c|c|c|}\hline
         &  \multicolumn{4}{|c|}{Plate with curvature at leading edge}&  \multicolumn{4}{|c|}{Plate without curvature at leading edge}\\\hline
         $\alpha$&  $Cl_{mean}$&  $Cl_{rms}$&  $Sk_{Cl}$&  $Ku_{Cl}$&  $Cl_{mean}$&  $Cl_{rms}$&  $Sk_{Cl}$& $Ku_{Cl}$\\\hline
         $0^\circ$&  0.132&  0.000&  -&  -&  0.002&  0.000&  -& -\\\hline
         $5^\circ$&  0.376&  0.024&  -0.020&  1.484
&  0.473&  0.000&  -& -\\\hline
         $10^\circ$&  0.609&  0.170&  0.172&  1.503
&  0.582&  0.081&  -0.224& 2.282
\\\hline
         $15^\circ$&  0.832&  0.185&  0.336&  1.524
&  0.895&  0.200&  0.257& 1.511
\\\hline
         $20^\circ$&  0.918&  0.211&  0.316&  1.997
&  0.867&  0.187&  0.353& 1.608
\\\hline
         $25^\circ$&  0.988&  0.262&  0.625&  2.057
&  0.972&  0.233&  0.507& 2.686\\ \hline
    \end{tabular}

\end{table*}

\begin{figure}
    \centering
    \includegraphics[width=1\linewidth]{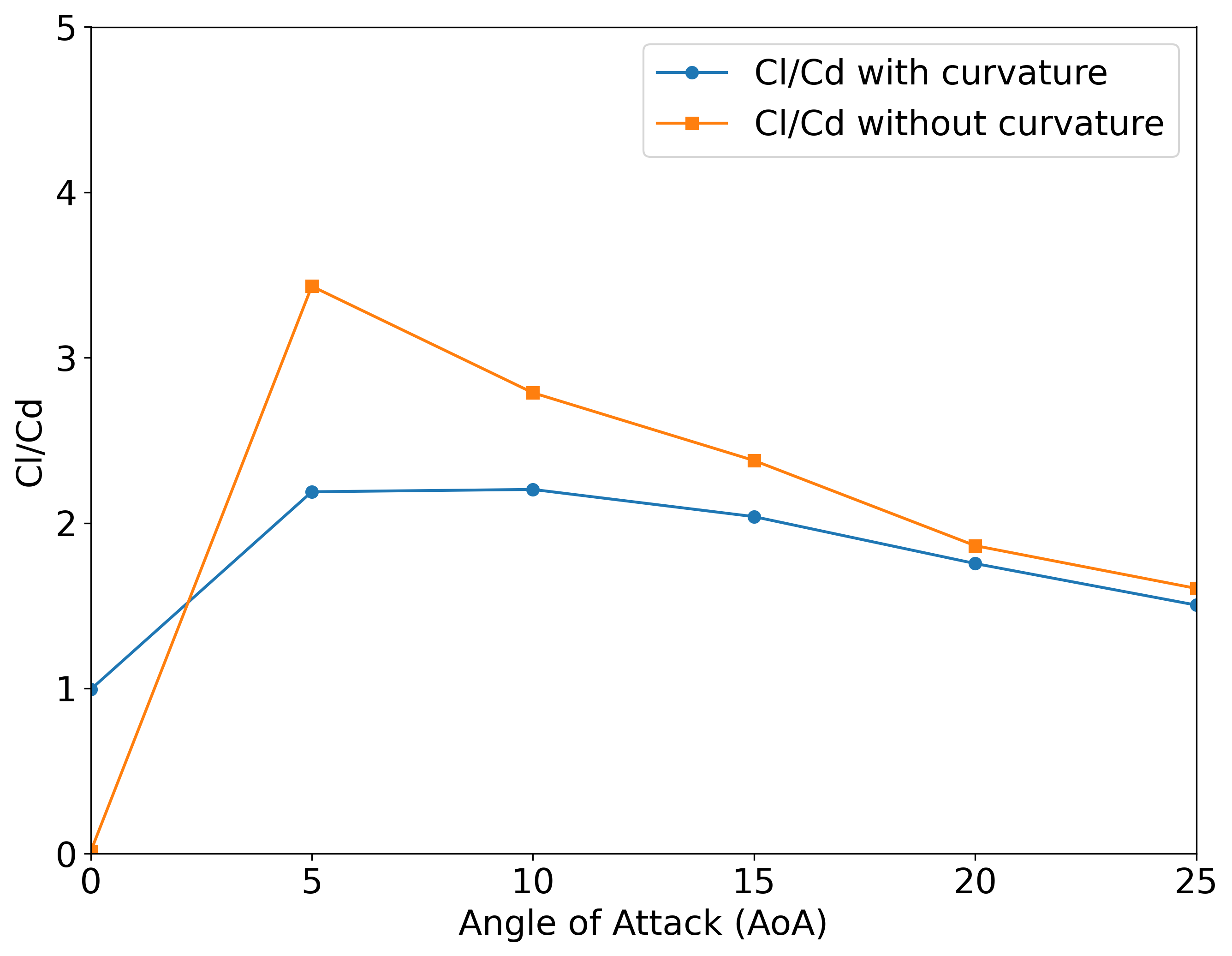}
    \caption{Glide ratio at various angle of attack (AoA)}
    \label{fig:glide_ratio}
\end{figure}

The combined effect of both lift and force coefficients for both the flat plate cases is shown in Fig. \ref{fig:glide_ratio}.  The plate with curvature at the leading edge has a value of $Cl/Cd \approx 1$ at $\alpha = 0^\circ$. For all other angle of attacks the plate without curvature has a higher glide ratio. The extreme cases like angle of attack $\alpha = 0^\circ$ and for $\alpha = 20^\circ$ and $25^\circ$ the plate with curvature competitive and comparable performance. The design information is relevant when designing fixed wing drone technologies were leading edge modification will help improve the overall performance of the fliers at low Reynolds number. 
\section{Discussions and Conclusion}
Modifying the leading edge for a blunt $90^\circ$ leading edge flat plate alters the fluid body interaction and influences the temporal variations in the aerodynamic loading. The present work uses numerical simulation to investigate the effect of simplified curvature at the leading edge of a flat plate at different angles of attack to the freestream flow at a low Re of 1000. The bluffness of the curvature initiates a flow separation at the leading edge which differs from the flow separated from the simple flat plate with thickness of 5\% of the length of the plate. The separated flow around the flat plate carries the signature of the geometry of the leading edge and the orientation of the flat plate. The flow around the flat plate with curvature has a thicker vorticity at lower angles of attack where the bluff body contribution to the vorticity is significant. Separated flow at the leading edge of simple flat plate reattaches at the plate surface at $\alpha = 5^\circ$ whereas the curvature at the leading edge initiates vortical regions within the reattachment region creating a virtually thicker shape and an early onset of shedding.  At higher angles of attack the wake is characterized by von-karman shedding with larger deflections for simple flat plate without curvature. The flat plate with curvature at the leading edge has larger suction side vortical structures.  

The shed vortices influence the temporal variation of the force coefficients. Multiple frequencies of $C_l$ time series noted at higher AOA ($\alpha = 20^\circ$) for the case with curvature and for $25^\circ$ for the flat plate without curvature.   The number of frequencies observed increases to two for AOA: $\alpha = 20^\circ$ and $25^\circ$ for flat plate with curvature . This multiple frequency shedding is observed for $\alpha = 25^\circ$ for simple flat plate.  The simple flat plate without curvature at the leading edge showed a quasiperiodic behavior for $\alpha = 10 ^\circ$.   The unsteadiness in the force coefficients  relatively similar for the drag coefficients but geometry  and orientation effect are more pronounced for the unsteadiness in the lift coefficients.  It is noted that the the flow $C_l$ skewness increases from negative to positive in the shedding region with a peak at $\alpha = 15^\circ$ for the case with curvature indicating the nature of vortical shedding. For varying angle of attacks there are no extreme excursions of the force coefficients and the force coefficients have flatter distribution about the mean.  Lift generation at $\alpha = 0^\circ$ for the plate with curvature at the leading edge gives it a space in the domain for flights at lower angle of attacks which is missing for the simple flat plate. At higher AOA the glide ratios are comparable for the plate with curvature at the leading edge and the simple flat plate. For the changing landscape of mobility, the information provided here will enable in making informed engineering design decisions for unmanned aerial or aquatic vehicles cruising at lower Reynolds numbers. 
This work opens up spaces for further studies on different geometries of protrusions/curvature at the leading edge at different Reynolds numbers.

\FloatBarrier
\section{References}
\bibliography{aipsamp}

\end{document}